\documentclass[letterpaper,english]{IEEEConf}
\usepackage[latin9]{inputenc}
\usepackage{geometry}
\usepackage{textcomp}
\usepackage{amsmath}
\usepackage{amsthm}
\usepackage{amssymb}
\usepackage{stackrel}
\usepackage{graphicx}
\usepackage{tablefootnote}

\makeatletter

\DeclareTextSymbolDefault{\textquotedbl}{T1}
\providecommand{\tabularnewline}{\\}

\theoremstyle{definition}
\newtheorem{assumption}{Assumption}
\theoremstyle{remark}
\newtheorem{rem}{\protect\remarkname}
\theoremstyle{plain}
\newtheorem{thm}{\protect\theoremname}

\IEEEoverridecommandlockouts
\usepackage{cite}
\usepackage{multirow}
\usepackage{tikz}
\usetikzlibrary{shapes.geometric, arrows, calc}
\usetikzlibrary{decorations.pathreplacing}
\usepackage[english]{babel}
\usepackage[compact]{titlesec}

\makeatother

\usepackage{babel}
\providecommand{\remarkname}{Remark}
\providecommand{\theoremname}{Theorem}

\begin{document}
\title{Online ResNet-Based Adaptive Control for Nonlinear Target Tracking}
\author{Cristian F. Nino, Omkar Sudhir Patil, Jordan C. Insinger, Marla R.
Eisman, and Warren E. Dixon\thanks{Cristian F. Nino, Omkar Sudhir Patil, Jordan C. Insinger, Marla R.
Eisman and Warren E. Dixon are with the Department of Mechanical and
Aerospace Engineering, University of Florida, Gainesville, FL, 32611,
USA. Email:\{cristian1928, patilomkarsudhir, jordan.insinger, marlaeisman,
wdixon\}@ufl.edu.}\thanks{This research is supported in part by AFOSR grant FA9550-19-1-0169,
AFRL grant FA8651-24-1-0018 , and AFOSR grant FA9550-21-1-0157. Any
opinions, findings and conclusions or recommendations expressed in
this material are those of the author(s) and do not necessarily reflect
the views of the sponsoring agency.}}
\maketitle
\begin{abstract}
A generalized ResNet architecture for adaptive control of nonlinear
systems with black box uncertainties is developed. The approach overcomes
limitations in existing methods by incorporating pre-activation shortcut
connections and a zeroth layer block that accommodates different input-output
dimensions. The developed Lyapunov-based adaptation law establishes
exponential convergence to a neighborhood of the target state despite
unknown dynamics and disturbances. Furthermore, the theoretical results
are validated through a comparative experiment.
\end{abstract}


\section{Introduction}

\IEEEPARstart{N}{eural} networks (NNs) are well established for approximating
unstructured uncertainties in continuous functions over compact domains
\cite{Stone1948,Hornik1991,Kidger.Lyons2020,Kratsios.Papon2022}.
The evolution of NN-based control has progressed from single-layer
architectures with Lyapunov-based adaptation \cite{Lewis1999,Lewis.Jagannathan.ea1998,Patre.Bhasin.ea2010}
to more complex deep neural network (DNN) implementations, motivated
by numerous examples of improved function approximation efficiency
\cite{LeCun.Bengio.ea2015,Goodfellow.Bengio.ea2016,Rolnick.Tegmark2018}.
Early DNN approaches develop Lyapunov-based adaptive update laws for
the output layer while the inner layers are updated either in an iterative
offline manner as in \cite{Joshi.Chowdhary2019} and \cite{Sun.Greene.ea2022},
or using modular adaptive update laws \cite{Le.Greene.ea2022}. Recent
developments have established frameworks for real-time adaptation
of all DNN layers \cite{Patil.Le.ea2022,Ryu.Choi2024} for various
DNN architectures, addressing issues in transient performance \cite{Le.Patil.ea2024}
and leveraging persistence of excitation \cite{Patil.Griffis.ea2023}.

Deep residual neural network (ResNet) architectures have emerged as
particularly promising candidates for adaptive control applications.
The ResNet architecture is popular because it addresses optimization
challenges that arise with increasing network depth, making them potentially
valuable for modeling complex system uncertainties.

The key innovation of ResNets is the introduction of skip connections
that create direct paths for information to flow through the network
during backpropagation as in \cite{He.Zhang.ea2016a} and \cite{Veit.Wilber.ea2016}.
These skip connections help prevent the degradation of gradient information
as it passes through multiple layers---a phenomenon known as the
vanishing gradient problem where the magnitude of gradients becomes
too small for effective weight updates in deep networks. Rather than
learning completely new representations at each layer, ResNets learn
the difference (or \textquotedbl residual\textquotedbl ) between
the input and the desired output of a layer, which simplifies the
optimization process. Theoretical analyses have demonstrated that
ResNets possess favorable optimization properties, including smoother
loss landscapes \cite{Li.Xu.ea2018}, absence of spurious local optima
with every local minimum being a global minimum as in \cite{Hardt.Ma2017}
and \cite{Kawaguchi.Bengio2019}, and stability of gradient descent
equilibria \cite{Nar.Sastry2018}. The universal approximation capabilities
of ResNets have also been investigated \cite{Lin.Jegelka2018,Tabuada.Gharesifard2023,Liu.Liang.ea2024},
confirming their ability to approximate any continuous function on
a compact set to arbitrary accuracy. A critical advancement in ResNet
design was the introduction of pre-activation shortcuts by \cite{He.Zhang.ea2016},
which position the skip connection before activation functions, improving
the flow of information through the network and enhancing the network's
ability to generalize to unseen data. This architectural modification
shares conceptual similarities with DenseNets \cite{Huang.Liu.ea2017},
which strengthen feature propagation through dense connectivity patterns
that connect each layer to every other layer, facilitating feature
reuse and enhancing information flow throughout the network.

Recently, \cite{Patil.Le.ea2022a} introduced the first Lyapunov-based
ResNet for online learning in control applications. However, their
implementation utilized the original ResNet architecture without incorporating
the pre-activation shortcut connections that have been demonstrated
to improve performance \cite{He.Zhang.ea2016}. This limitation potentially
restricts the learning capabilities and convergence properties of
their approach. Additionally, this approach requires the inputs and
outputs of the ResNet architecture to be of the same dimensions, thus
limiting the applicability of the development. These limitations potentially
restrict the learning capabilities and convergence properties of their
approach.

This work presents a generalized ResNet architecture featuring pre-activation
shortcut connections and a zeroth layer block designed for target
tracking of nonlinear systems. The developed approach positions the
skip connection before activation, with post-activation feeding into
the DNN block, leveraging the improved information propagation through
the network established by \cite{He.Zhang.ea2016}. Additionally,
the zeroth layer block can compensate for uncertainties with different
input and output size, overcoming the limitations of the approaches
in \cite{Patil.Le.ea2022a}. Similar to DenseNets \cite{Huang.Liu.ea2017},
the developed architecture facilitates stronger feature propagation
and reuse, and is specifically designed for online learning in control
applications. The key contribution of this work is the development
and analysis of a Lyapunov-based adaptation law for this generalized
ResNet architecture, which establishes exponential convergence to
a neighborhood of the target state despite unknown dynamics and disturbances.
Furthermore, the theoretical results are validated through a comparative
experiment.

\section{Deep Residual Neural Network Model}

Consider a fully connected feedforward ResNet with $b\in\mathbb{Z}_{\geq0}$
building blocks, input $x\in\mathbb{R}^{L_{\text{in}}}$, and output
$y\in\mathbb{R}^{L_{\text{out}}}$. For each block index $i\in\left\{ 0,\ldots,b\right\} $,
let $k_{i}\in\mathbb{Z}_{>0}$ be the number of hidden layers in the
$i^{\text{th}}$ block, let $\kappa_{i}\in\mathbb{R}^{L_{i,0}}$ denote
the block input (with $\kappa_{0}\triangleq x$ and $L_{0,0}\triangleq L_{\text{in}}$),
and let $\theta_{i}\in\mathbb{R}^{p_{i}}$ be the vector of parameters
(weights and biases) associated with the $i^{\text{th}}$ block.

For each block $i\in\left\{ 0,\ldots,b\right\} $, let $L_{i,j}\in\mathbb{Z}_{>0}$
denote the number of neurons in the $j^{\text{th}}$ layer for $j\in\left\{ 0,\ldots,k_{i}+1\right\} $.
Furthermore, define the augmented dimension $L_{i,j}^{a}\triangleq L_{i,j}+1$,
for all $\left(i,j\right)\in\left\{ 0,\ldots,b\right\} \times\left\{ 0,\ldots,k_{i}\right\} $.
Each block function $\Phi_{i}:\mathbb{R}^{L_{i,0}^{a}}\times\mathbb{R}^{p_{i}}\to\mathbb{R}^{L_{i,k_{i}+1}}$
is a fully connected feedforward DNN, with $L_{i,k_{i}+1}\triangleq L_{\text{out}}$
for all $\left(i,j\right)\in\left\{ 0,\ldots,b\right\} \times\left\{ 0,\ldots,k_{i}\right\} $.
For any input $v\in\mathbb{R}^{L_{i,j}^{a}}$, the DNN is defined
recursively by

\begin{equation}
\varphi_{i,j}\left(v\right)\triangleq\begin{cases}
V_{i,0}^{\top}v, & j=0,\\
V_{i,j}^{\top}\phi_{i,j}\left(\varphi_{i,j-1}\left(v\right)\right), & j\in\left\{ 1,\ldots,k_{i}\right\} ,
\end{cases}\label{eq:DNN}
\end{equation}
with $\Phi_{i}\left(v,\theta_{i}\right)=\varphi_{i,k_{i}}\left(v\right)$. 

For each $j\in\left\{ 0,1,\ldots,k_{i}\right\} $ the matrix $V_{i,j}\in\mathbb{R}^{L_{i,j}^{a}\times L_{i,j+1}}$
contains the weights and biases; in particular, if a layer has $n$
(augmented) inputs and the subsequent layer has $m$ nodes, then $V\in\mathbb{R}^{n\times m}$
is constructed so that its $\left(i,j\right)^{\text{th}}$ entry represents
the weight from the $i^{\text{th}}$ node of the input to the $j^{\text{th}}$
node of the output, with the last row corresponding to the bias terms.
For the DNN architecture described by (\ref{eq:DNN}), the vector
of DNN weights of the $i^{\text{th}}$ block is $\theta_{i}\triangleq\left[\begin{array}{ccc}
\text{vec}\left(V_{i,0}\right)^{\top} & \cdots & \text{vec}\left(V_{i,k_{i}}\right)^{\top}\end{array}\right]^{\top}\in\mathbb{R}^{p_{i}}$, where $p_{i}\triangleq\sum_{j=0}^{k_{i}}L_{i,j}^{a}L_{i,j+1}$,
and $\text{vec}\left(V_{i,j}\right)$ denotes the vectorization of
$V_{i,j}$ performed in column-major order (i.e., the columns are
stacked sequentially to form a vector). The activation function $\phi_{i,j}:\mathbb{R}^{L_{i,j}}\to\mathbb{R}^{L_{i,j}^{a}}$
is given by $\phi_{i,j}\left(\varphi_{i,j-1}\right)$ $=$ $\left[\varsigma_{i,1}\left(\left(\varphi_{i,j-1}\right)_{1}\right)\right.$
$\begin{array}{c}
\varsigma_{i,2}\left(\left(\varphi_{i,j-1}\right)_{2}\right)\end{array}$ $\begin{array}{c}
\cdots\end{array}$ $\begin{array}{c}
\varsigma_{i,L_{i,j}}\left(\left(\varphi_{i,j-1}\right)_{L_{i,j}}\right)\end{array}$ $\left.1\right]^{\top}$ $\in\mathbb{R}^{L_{i,j}^{a}}$ where $\left(\varphi_{i,j-1}\right)_{\ell}$
denotes the $\ell^{\text{th}}$ component of $\varphi_{i,j-1}$, each
$\varsigma_{i,j}:\mathbb{R}\to\mathbb{R}$ denotes a smooth activation
function, and 1 denotes the augmented hidden layer that accounts for
the bias term.

\begin{figure}
\begin{centering}
\includegraphics[width=1\columnwidth]{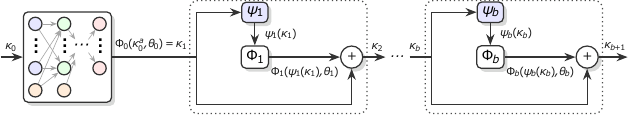}
\par\end{centering}
\caption{Deep Residual Neural Network Architecture}
\end{figure}
A pre-activation design is used so that, before each block (except
block $0$), the output of the previous block is processed by an external
activation function. Specifically, for each block $i\in\left\{ 1,\ldots,b\right\} $,
define the pre-activation mapping $\psi_{i}:\mathbb{R}^{L_{i,k_{i}+1}}\to\mathbb{R}^{L_{i,k_{i}+1}^{a}}$
by $\psi_{i}\left(\kappa_{i}\right)$ $=$ $\left[\varrho_{i,1}\left(\left(\kappa_{i}\right)_{1}\right)\right.$
$\varrho_{i,2}\left(\left(\kappa_{i}\right)_{2}\right)$ $\begin{array}{c}
\cdots\end{array}$ $\varrho_{i,L_{i,k_{i}+1}}\left(\left(\kappa_{i}\right)_{L_{i,k_{i}+1}}\right)$
$\left.1\right]^{\top}$ $\in\mathbb{R}^{L_{i,k_{i}+1}^{a}}$ where
$\left(\kappa_{i}\right)_{\ell}$ denotes the $\ell^{\text{th}}$
component of $\kappa_{i}$, each $\varrho_{i,j+1}:\mathbb{R}\to\mathbb{R}$
denotes a smooth activation function, and 1, again, accounts for the
bias term. The output of $\psi_{i}$ serves as the input to block
$i$ and the residual connection is implemented by adding the current
block output to the pre-activated output from the previous block.
Hence, the ResNet recursion is defined by
\begin{equation}
\kappa_{i+1}\triangleq\begin{cases}
\Phi_{0}\left(\kappa_{0}^{a},\theta_{0}\right), & i=0,\\
\kappa_{i}+\Phi_{i}\left(\psi_{i}\left(\kappa_{i}\right),\theta_{i}\right), & i\in\left\{ 1,\ldots,b\right\} ,
\end{cases}\label{eq:ResNet}
\end{equation}
with output $y\in\mathbb{R}^{L_{\text{out}}}$ and overall parameter
vector $\Theta\triangleq\left[\begin{array}{ccc}
\theta_{0}^{\top} & \cdots & \theta_{b}^{\top}\end{array}\right]^{\top}\in\mathbb{R}^{{\tt p}}$, with ${\tt p}\triangleq\sum_{i=0}^{b}p_{i}$, where $\kappa_{0}^{a}\triangleq\left[\begin{array}{cc}
\kappa_{0}^{\top} & 1\end{array}\right]^{\top}\in\mathbb{R}^{L_{0,0}^{a}}$ denotes the augmented input to block $i=0$. Therefore, the complete
ResNet is represented as $\Psi:\mathbb{R}^{L_{\text{in}}}\times\mathbb{R}^{{\tt p}}\to\mathbb{R}^{L_{\text{out}}}$
expressed as $\Psi\left(\kappa,\Theta\right)=\kappa_{b+1}$.

The Jacobian of the ResNet is represented as $\frac{\partial}{\partial\Theta}\Psi\left(\kappa,\Theta\right)=\left[\begin{array}{ccc}
\frac{\partial}{\partial\theta_{0}}\Psi\left(\kappa,\Theta\right) & \cdots & \frac{\partial}{\partial\theta_{b}}\Psi\left(\kappa,\Theta\right)\end{array}\right]\in\mathbb{R}^{L_{\text{out}}\times{\tt p}}$ and $\frac{\partial}{\partial\theta_{i}}\Psi\left(\kappa,\Theta\right)=\left[\begin{array}{ccc}
\frac{\partial}{\partial{\rm vec}\left(V_{i,0}\right)}\Psi\left(\kappa,\Theta\right) & \cdots & \frac{\partial}{\partial{\rm vec}\left(V_{i,k_{i}}\right)}\Psi\left(\kappa,\Theta\right)\end{array}\right]\in\mathbb{R}^{L_{\text{out}}\times p_{i}}$, where $\frac{\partial}{\partial{\rm vec}\left(V_{i,j}\right)}\Psi\left(\kappa,\Theta\right)\in\mathbb{R}^{L_{\text{out}}\times L_{i,j}^{a}L_{i,j+1}}$
for all $j\in\left\{ 0,\ldots,k_{i}\right\} $. Using (\ref{eq:DNN}),
(\ref{eq:ResNet}), and the property of the vectorization operator
yields{\scriptsize{}
\begin{align*}
\frac{\partial\Psi}{\partial\mathrm{vec}\left(V_{i,j}\right)} & =\left(\stackrel{\curvearrowleft}{\stackrel[m=i+1]{b}{\prod}}\left(I_{L_{\text{out}}}+\left(\stackrel{\curvearrowleft}{\stackrel[\ell=1]{k_{m}}{\prod}}V_{m,\ell}^{\top}\frac{\partial\phi_{m,\ell}}{\partial\varphi_{m,\ell-1}}\right)V_{m,0}^{\top}\frac{\partial\psi_{m}}{\partial\kappa_{m}}\right)\right)\\
 & \hspace{1em}\cdot\left(\stackrel{\curvearrowleft}{\stackrel[\ell=j+1]{k_{i}}{\prod}}V_{i,\ell}^{\top}\frac{\partial\phi_{i,\ell}}{\partial\varphi_{i,\ell-1}}\right)\left(I_{L_{i,j+1}}\otimes\varkappa_{i,j}^{\top}\right),
\end{align*}
}where $\varkappa_{i,j}\triangleq\kappa_{0}^{a}$ if $i=0$ and $j=0$,
$\phi_{0,j}\left(\varphi_{0,j-1}\left(\kappa_{0}^{a}\right)\right)$
if $i=0$ and $j>0$, $\psi_{i}\left(\kappa_{i}\right)$ if $i>0$
and $j=0$, $\phi_{i,j}\left(\varphi_{i,j-1}\left(\psi_{i}\left(\kappa_{i}\right)\right)\right)$
if $i>0$ and $j>0$.

The Jacobian $\frac{\partial\phi_{i,j}\left(\varphi_{i,j-1}\left(v\right)\right)}{\partial\varphi_{i,j-1}\left(v\right)}:\mathbb{R}^{L_{i,j}}\to\mathbb{R}^{L_{i,j}^{a}\times L_{i,j}}$
of the activation function vector at the $j^{\mathrm{th}}$ layer
is given by $\frac{\partial\phi_{i,j}\left(\varphi_{i,j-1}\left(v\right)\right)}{\partial\varphi_{i,j-1}\left(v\right)}$
$=$ $\left[\text{diag}\left\{ \frac{{\rm d}\varsigma_{i,1}\left(\left(\varphi_{i,j-1}\right)_{1}\right)}{{\rm d}\left(\varphi_{i,j-1}\right)_{1}},\ldots,\frac{{\rm d}\varsigma_{i,L_{i,j}}\left(\left(\varphi_{i,j-1}\right)_{L_{i,j}}\right)}{{\rm d}\left(\varphi_{i,j-1}\right)_{L_{i,j}}}\right\} \right.$$\left.\mathbf{0}_{L_{i,j}}^{\top}\right]^{\top}$.
Similarly, the Jacobian $\frac{\partial\psi_{m}\left(\kappa_{m}\right)}{\partial\kappa_{m}}:\mathbb{R}^{L_{\text{out}}}\to\mathbb{R}^{L_{\text{out}}^{a}\times L_{\text{out}}}$
of the pre-activation function vector at block $m$ is given by $\frac{\partial\psi_{m}\left(\kappa_{m}\right)}{\partial\kappa_{m}}$
$=$ $\left[\text{diag}\left\{ \frac{{\rm d}\varrho_{m,1}\left(\left(\kappa_{m}\right)_{1}\right)}{{\rm d}\left(\kappa_{m}\right)_{1}},\ldots,\frac{{\rm d}\varrho_{m,L_{\text{out}}}\left(\left(\kappa_{m}\right)_{L_{\text{out}}}\right)}{{\rm d}\left(\kappa_{m}\right)_{L_{\text{out}}}}\right\} \right.$$\left.\mathbf{0}_{L_{\text{out}}}^{\top}\right]^{\top}$.

\section{Problem Formulation}

Consider the second-order nonlinear dynamical system described by
the differential equation
\begin{equation}
\ddot{q}=f\left(q,\dot{q}\right)+g\left(q,\dot{q},t\right)u+\omega\left(t\right),\label{eq:Agent Dynamics}
\end{equation}
where $t\in\mathbb{R}_{\geq0}$ denotes time, $q\in\mathbb{R}^{n}$
represents the generalized position, $\dot{q}\in\mathbb{R}^{n}$ the
generalized velocity, $\ddot{q}\in\mathbb{R}^{n}$ the generalized
acceleration, $f\in\mathcal{C}^{1}\left(\mathbb{R}^{n}\times\mathbb{R}^{n};\mathbb{R}^{n}\right)$
represents unknown system dynamics, $g\in\mathcal{C}^{1}\left(\mathbb{R}^{n}\times\mathbb{R}^{n}\times\mathbb{R}_{\geq0};\mathbb{R}^{n\times m}\right)$
denotes the known control effectiveness matrix, $\omega\in\mathcal{C}^{1}\left(\mathbb{R}_{\geq0};\mathbb{R}^{n}\right)$
represents an exogenous disturbance, and $u:\mathbb{R}_{\geq0}\to\mathbb{R}^{m}$
denotes the control input signal.

The following assumptions and properties hold. First, the matrix $g\left(q,\dot{q},t\right)$
has full row-rank for all $\left(q,\dot{q},t\right)\in\mathbb{R}^{n}\times\mathbb{R}^{n}\times\mathbb{R}_{\geq0}$.
Second, the mapping $t\mapsto g\left(q,\dot{q},t\right)$ is uniformly
bounded for all states $\left(q,\dot{q}\right)\in\mathbb{R}^{n}\times\mathbb{R}^{n}$.
Third, there exists a known constant $\overline{\omega}\in\mathbb{R}_{\geq0}$
such that $\left\Vert \omega\left(t\right)\right\Vert \leq\overline{\omega}$
for all $t\in\mathbb{R}_{\geq0}$.

By the full row-rank property of $g$, its right Moore-Penrose pseudoinverse
$g^{+}\in\mathcal{C}^{1}\left(\mathbb{R}^{n}\times\mathbb{R}^{n}\times\mathbb{R}_{\geq0};\mathbb{R}^{m\times n}\right)$
exists, defined as $g^{+}\left(q,\dot{q},t\right)\triangleq g^{\top}\left(q,\dot{q},t\right)\left(g\left(q,\dot{q},t\right)g^{\top}\left(q,\dot{q},t\right)\right)^{-1}$.
Furthermore, the mapping $t\mapsto g^{+}\left(q,\dot{q},t\right)$
is uniformly bounded for all states $\left(q,\dot{q}\right)\in\mathbb{R}^{n}\times\mathbb{R}^{n}$.

The reference trajectory is governed by the autonomous second-order
system
\begin{equation}
\ddot{q}_{d}=f_{d}\left(q_{d},\dot{q}_{d}\right),\label{eq:Target Dynamics}
\end{equation}
where $q_{d}\in\mathbb{R}^{n}$ denotes the reference position, $\dot{q}_{d}\in\mathbb{R}^{n}$
denotes the reference velocity, $\ddot{q}_{d}\in\mathbb{R}^{n}$ denotes
the reference acceleration, and $f_{d}:\mathbb{R}^{n}\times\mathbb{R}^{n}\to\mathbb{R}^{n}$
represents the unknown reference dynamics.
\begin{assumption}
\label{targetbounds}There exist known constants $\overline{q}_{d},\overline{\dot{q}}_{d}\in\mathbb{R}_{>0}$
such that $q_{d},\dot{q}_{d}\in\mathcal{L}_{\infty}\left(\mathbb{R}_{\geq0};\mathbb{R}^{n}\right)$.
\end{assumption}
The control objective is to design a ResNet-based adaptive controller
such that the state trajectory $q$ is exponentially regulated to
a neighborhood of the reference trajectory $q_{d}$, despite the presence
of unknown dynamics and bounded disturbances. To facilitate the control
objective, define the tracking error $e\in\mathbb{R}^{n}$ as
\begin{equation}
e\triangleq q_{d}-q.\label{eq:Tracking Error}
\end{equation}

\section{Control Synthesis}

To facilitate the control design, the auxiliary tracking error function
$r\in\mathbb{R}^{n}$ is defined as
\begin{equation}
r\triangleq\dot{e}+k_{1}e,\label{eq: Filtered Tracking Error}
\end{equation}
where $k_{1}\in\mathbb{R}_{>0}$ is a constant control gain. Differentiating
(\ref{eq: Filtered Tracking Error}) with respect to time and substituting
(\ref{eq:Agent Dynamics})-(\ref{eq: Filtered Tracking Error}) yields{\small{}
\begin{equation}
\begin{aligned}\dot{r} & =h\left(q,\dot{q},q_{d},\dot{q}_{d}\right)-g\left(q,\dot{q},t\right)u-\omega\left(t\right)+k_{1}\left(r-k_{1}e\right),\end{aligned}
\label{eq:rDot}
\end{equation}
}where $h:\mathbb{R}^{n}\times\mathbb{R}^{n}\times\mathbb{R}^{n}\times\mathbb{R}^{n}\to\mathbb{R}^{n}$
is defined as $h\left(q,\dot{q},q_{d},\dot{q}_{d}\right)\triangleq f_{d}\left(q_{d},\dot{q}_{d}\right)-f\left(q,\dot{q}\right)$.

\subsection{Residual Neural Network Function Approximation}

The ResNet architecture, characterized by skip connections and hierarchical
feature extraction, models incremental changes rather than complete
transformations of the underlying nonlinear mapping. This architecture
learns the differences (or \textquotedbl residuals\textquotedbl )
between input and desired output at each layer, thereby enabling effective
function approximation for complex nonlinear systems without requiring
explicit governing equations.

To approximate the unknown dynamics given by $h$ in (\ref{eq:rDot}),
define the input vector $\kappa:\mathbb{R}_{\geq0}\to\mathbb{R}^{4n}$
as $\kappa\triangleq\left[\begin{array}{cccc}
q^{\top} & \dot{q}^{\top} & q_{d}^{\top} & \dot{q}_{d}^{\top}\end{array}\right]^{\top}\in\Omega$, where $\Omega\subset\mathbb{R}^{4n}$ is a compact set over which
the universal approximation property holds. The ResNet-based approximation
of $h\left(\kappa\right)$ is given by $\Psi\left(\kappa,\widehat{\Theta}\right)$,
where $\Psi:\mathbb{R}^{4n}\times\mathbb{R}^{{\tt p}}\to\mathbb{R}^{n}$
denotes the ResNet architecture mapping and $\widehat{\Theta}\in\mathbb{R}^{{\tt p}}$
denotes the adaptive parameter estimate.

The approximation objective is to determine optimal estimates $\widehat{\Theta}$
within a predefined search space such that the mapping $\kappa\mapsto\Psi\left(\kappa,\widehat{\Theta}\right)$
approximates $\kappa\mapsto h\left(\kappa\right)$ with minimal error
for all $\kappa\in\Omega$. Let $\mho\subset\mathbb{R}^{{\tt p}}$
denote a user-selected compact, convex parameter search space with
a $\mathcal{C}^{\infty}$ boundary, satisfying $\mathbf{0}_{{\tt p}}\in\text{int}\left(\mho\right)$,
and define $\overline{\Theta}\triangleq\underset{\Theta\in\mho}{\max}\left\Vert \Theta\right\Vert .$
An objective function $\mathcal{J}:\mho\to\mathbb{R}_{\geq0}$ is
selected to quantify the quality of the approximation achieved by
the parameters $\Theta\in\mho$, where $\Theta\in\mathbb{R}^{{\tt p}}$
is an arbitrary parameter vector.\footnote{The function $\mathcal{J}\left(\Theta\right)$ typically incorporates
a measure of the approximation error, often based on a norm of the
difference $h\left(\kappa\right)-\Psi\left(\kappa,\Theta\right)$
aggregated over the domain $\Omega$. A common example is the mean
squared error functional, $\mathcal{J}_{\text{MSE}}\left(\Theta\right)\triangleq\int_{\Omega}\left\Vert h\left(\kappa\right)-\Psi\left(\kappa,\Theta\right)\right\Vert ^{2}{\rm d}\mu\left(\kappa\right)$,
where $\mu$ is a suitable measure on $\Omega$. However, the specific
choice of $\mathcal{J}$ can be adapted based on the application requirements
or available data.}
\begin{assumption}
\label{thm:The-loss-function}The selected objective function $\mathcal{J}:\mho\to\mathbb{R}_{\geq0}$
is continuous and strictly convex on the compact, convex set $\mho$.
\end{assumption}
Under Assumption \ref{thm:The-loss-function}, the existence and uniqueness
of a global minimizer for $\mathcal{J}$ within the search space $\mho$
is guaranteed. This unique optimal parameter vector within the search
space is denoted by $\Theta^{\ast}\in\mho$, defined as\footnote{For ResNets with arbitrary nonlinear activation functions, it has
been established that every local minimum is a global minimum \cite{Kawaguchi.Bengio2019}.}
\begin{align}
\Theta^{\ast} & \triangleq\underset{\Theta\in\mho}{\arg\min\ }\mathcal{J}\left(\Theta\right).\label{eq:theta star}
\end{align}

\begin{rem}
\label{rem:Universal Function Approximation} The universal function
approximation property of ResNets was not invoked in the definition
of $\Theta^{\ast}$. The universal function approximation theorem
states that the function space of ResNets is dense in the space of
continuous functions $\mathcal{C}\left(\Omega\right)$ \cite{Lin.Jegelka2018,Liu.Liang.ea2024}.
Consequently, for any prescribed $\overline{\varepsilon}>0$, there
exists a ResNet $\Psi$ and a corresponding parameter vector $\Theta$
such that $\underset{\kappa\in\Omega}{\sup}\left\Vert h\left(\kappa\right)-\Psi\left(\kappa,\Theta\right)\right\Vert <\overline{\varepsilon}$.
This implies $\int_{\Omega}\left\Vert h\left(\kappa\right)-\Psi\left(\kappa,\Theta\right)\right\Vert ^{2}{\rm d}\mu\left(\kappa\right)<\overline{\varepsilon}^{2}\mu\left(\Omega\right)$.
However, determining a search space $\mho$ for arbitrary $\overline{\varepsilon}$
remains an open challenge. Therefore, $\mho$ is arbitrarily selected,
at the expense of guarantees on the approximation accuracy.
\end{rem}

\subsection{Control Design}

Following the previous discussion, the unknown dynamics $h\left(\kappa\right)$
in (\ref{eq:rDot}) are modeled using a ResNet as

\begin{equation}
h\left(\kappa\right)=\Psi\left(\kappa,\Theta^{\ast}\right)+\varepsilon\left(\kappa\right),\label{eq:UFAP}
\end{equation}
where $\varepsilon:\mathbb{R}^{4n}\to\mathbb{R}^{n}$ is an unknown
function representing the optimal reconstruction error associated
with $\Theta^{\ast}$, which is bounded as
\begin{equation}
\sup_{\kappa\in\Omega}\left\Vert \varepsilon\left(\kappa\right)\right\Vert <\overline{\varepsilon}.\label{eq: reconstruction error bound}
\end{equation}
Based on (\ref{eq:rDot}) and the subsequent stability analysis, the
control input is designed as{\small{}
\begin{equation}
\begin{aligned}u & =g^{+}\left(q,\dot{q},t\right)\left(\left(1-k_{1}^{2}\right)e+\left(k_{1}+k_{2}\right)r+\Psi\left(\kappa,\widehat{\Theta}\right)\right)\end{aligned}
,\label{eq:control input}
\end{equation}
}where $k_{2}\in\mathbb{R}_{>0}$ is a constant control gain. Substituting
(\ref{eq:UFAP}) and (\ref{eq:control input}) into (\ref{eq:rDot})
yields{\small{}
\begin{equation}
\begin{aligned}\dot{r} & =-k_{2}r-e+\varepsilon\left(\kappa\right)-\omega\left(t\right)+\Psi\left(\kappa,\Theta^{\ast}\right)-\Psi\left(\kappa,\widehat{\Theta}\right).\end{aligned}
\label{eq:rDot-1}
\end{equation}
}Based on the subsequent stability analysis, the adaptive update law
for $\widehat{\Theta}$ is designed as
\begin{equation}
\dot{\widehat{\Theta}}=\text{proj}_{\mho}\left(\Gamma\left(\frac{\partial\Psi\left(\kappa,\widehat{\Theta}\right)}{\partial\widehat{\Theta}}^{\top}r-k_{3}\widehat{\Theta}\right)\right),\label{eq:Update Law}
\end{equation}
where $k_{3}\in\mathbb{R}_{>0}$ is a constant control gain, $\Gamma\in\mathbb{R}^{{\tt p}\times{\tt p}}$
is a user-defined positive-definite learning rate matrix, and the
projection operator ensures $\widehat{\Theta}\in\mho$, defined as
in \cite[Appendix E]{Krstic.Kanellakopoulos.ea1995}.

\section{Stability Analysis}

The ResNet mapping $\Psi\left(\kappa,\Theta\right)$ described in
(\ref{eq:UFAP}) is inherently nonlinear with respect to its weights.
Designing adaptive controllers and performing stability analyses for
systems that are nonlinearly parameterizable presents significant
theoretical challenges. A method to address the nonlinear structure
of the uncertainty, especially for ResNets, is to use a first-order
Taylor series approximation. To quantify the approximation, the parameter
estimation error $\widetilde{\Theta}\in\mathbb{R}^{{\tt p}}$ is defined
as
\begin{equation}
\widetilde{\Theta}=\Theta^{\ast}-\widehat{\Theta}.\label{eq: parameter estimation error}
\end{equation}
Applying first-order Taylor's theorem to the mapping $\Theta\mapsto\Psi\left(\kappa,\Theta\right)$
evaluated at $\widehat{\Theta}$, and using (\ref{eq: parameter estimation error})
yields
\begin{equation}
\Psi\left(\kappa,\Theta^{\ast}\right)-\Psi\left(\kappa,\widehat{\Theta}\right)=\frac{\partial\Psi\left(\kappa,\widehat{\Theta}\right)}{\partial\widehat{\Theta}}\widetilde{\Theta}+R\left(\kappa,\widetilde{\Theta}\right),\label{eq: Taylors Theorem}
\end{equation}
where $R:\mathbb{R}^{4n}\times\mathbb{R}^{{\tt p}}\to\mathbb{R}^{n}$
denotes the Lagrange remainder term. Substituting (\ref{eq: Taylors Theorem})
into (\ref{eq:rDot-1}) yields{\footnotesize{}
\begin{equation}
\begin{aligned}\dot{r} & =\frac{\partial\Psi\left(\kappa,\widehat{\Theta}\right)}{\partial\widehat{\Theta}}\widetilde{\Theta}+R\left(\kappa,\widetilde{\Theta}\right)+\varepsilon\left(\kappa\right)-k_{2}r-e-\omega\left(t\right).\end{aligned}
\label{eq:rDot-2}
\end{equation}
}{\footnotesize\par}

Let $z\in\mathbb{R}^{\varphi}$ denote the concatenated state vector
$z\triangleq\left[\begin{array}{ccc}
e^{\top} & r^{\top} & \widetilde{\Theta}^{\top}\end{array}\right]^{\top}$, where $\varphi\triangleq2n+{\tt p}$. The evolution of $z$ is governed
by the initial value problem
\begin{equation}
\dot{z}={\tt f}\left(z,t\right),\ z\left(t_{0}\right)=z_{0},\label{eq:IVP}
\end{equation}
where, $t_{0}\geq0$ is the initial time, $z_{0}\in\mathbb{R}^{\varphi}$
is the initial state, and, using (\ref{eq:Tracking Error}), (\ref{eq:Update Law}),
(\ref{eq: parameter estimation error}), and (\ref{eq:rDot-2}), the
vector field ${\tt f}:\mathbb{R}^{\varphi}\times\mathbb{R}_{\geq0}\to\mathbb{R}^{\varphi}$
is defined as
\begin{equation}
{\tt f}\left(z,t\right)=\left[\begin{array}{c}
r-k_{1}e\\
\left(\begin{array}{c}
\frac{\partial\Psi\left(\kappa,\widehat{\Theta}\right)}{\partial\widehat{\Theta}}\widetilde{\Theta}+R\left(\kappa,\widetilde{\Theta}\right)\\
+\varepsilon\left(\kappa\right)-k_{2}r-e-\omega\left(t\right).
\end{array}\right)\\
-\text{proj}_{\mho}\left(\Gamma\left(\frac{\partial\Psi\left(\kappa,\widehat{\Theta}\right)}{\partial\widehat{\Theta}}^{\top}r-k_{3}\widehat{\Theta}\right)\right)
\end{array}\right].\label{eq:CLES}
\end{equation}

Since the universal approximation property of the ResNet holds only
on the compact domain $\Omega$, the subsequent stability analysis
requires ensuring $\kappa\in\Omega$. This is achieved by establishing
a stability result which constrains the solution $z$ to a compact
domain. Consider the Lyapunov function candidate $V:\mathbb{R}^{\varphi}\to\mathbb{R}_{\geq0}$
defined as
\begin{equation}
V\left(z\right)\triangleq\frac{1}{2}z^{\top}Pz,\label{eq:Lyapunov}
\end{equation}
where $P\triangleq\text{blkdiag}\left\{ I_{2n},\Gamma^{-1}\right\} \in\mathbb{R}^{\varphi\times\varphi}$.
By the Rayleigh quotient, (\ref{eq:Lyapunov}) satisfies
\begin{equation}
\frac{1}{2}\lambda_{1}\left\Vert z\right\Vert ^{2}\leq V\left(z\right)\leq\frac{1}{2}\lambda_{\varphi}\left\Vert z\right\Vert ^{2},\label{eq: RQ}
\end{equation}
where $\lambda_{1}\triangleq\lambda_{\min}\left\{ P\right\} =\min\left\{ 1,\lambda_{\min}\left\{ \Gamma^{-1}\right\} \right\} $
and $\lambda_{\varphi}\triangleq\lambda_{\max}\left\{ P\right\} =\max\left\{ 1,\lambda_{\max}\left\{ \Gamma^{-1}\right\} \right\} $.
Based on the subsequent analysis, define $\delta\triangleq\frac{3\left(\overline{\omega}+\overline{\varepsilon}\right)^{2}}{4k_{2}}+\frac{k_{3}\overline{\Theta}^{2}}{2}$
and $k_{\min}\triangleq\min\left\{ k_{1},\frac{1}{3}k_{2},\frac{1}{2}k_{3}\right\} $.
Furthermore, let $\rho:\mathbb{R}_{\geq0}\to\mathbb{R}_{\geq0}$ denote
a strictly increasing function that is subsequently defined, and define
$\overline{\rho}\left(\cdot\right)\triangleq\rho\left(\cdot\right)-\rho\left(0\right)$,
where $\overline{\rho}$ is strictly increasing and invertible. The
region to which the state trajectory is constrained is defined as{\small{}
\begin{equation}
\mathcal{D}\triangleq\left\{ \iota\in\mathbb{R}^{\varphi}:\left\Vert \iota\right\Vert \leq\overline{\rho}^{-1}\left(k_{2}\left(k_{\min}-\lambda_{V}\right)-\rho\left(0\right)\right)\right\} ,\label{eq:state space}
\end{equation}
}where $\lambda_{V}\in\mathbb{R}_{>0}$ is a user-defined rate of
convergence parameter. The compact domain $\Omega\subset\mathbb{R}^{4n}$
over which the universal approximation property must hold is selected
as
\begin{equation}
\begin{aligned}\Omega & \triangleq\left\{ \iota\in\mathbb{R}^{4n}\right.:\left\Vert \iota\right\Vert \leq2\left(\overline{q}_{d}+\overline{\dot{q}}_{d}\right)\\
 & +\left(k_{1}+2\right)\overline{\rho}^{-1}\left.\left(k_{2}\left(k_{\min}-\lambda_{V}\right)-\rho\left(0\right)\right)\right\} .
\end{aligned}
\label{eq:OmegaSet}
\end{equation}
For the dynamical system described by (\ref{eq:IVP}), the set of
initial conditions $\mathcal{S}\subset\mathcal{D}$ is defined as
\begin{equation}
\begin{aligned}\mathcal{S} & \triangleq\left\{ \iota\in\mathbb{R}^{\varphi}\right.:\left\Vert \iota\right\Vert \leq-\sqrt{\frac{\delta}{\lambda_{V}}}\\
 & +\sqrt{\frac{\lambda_{1}}{\lambda_{\varphi}}}\left.\overline{\rho}^{-1}\left(k_{2}\left(k_{\min}-\lambda_{V}\right)-\rho\left(0\right)\right)\right\} ,
\end{aligned}
\label{eq:initial conditions}
\end{equation}
and the uniform ultimate bound $\mathcal{U}\subset\mathbb{R}^{\varphi}$
is defined as
\begin{equation}
\mathcal{U}\triangleq\left\{ \iota\in\mathbb{R}^{\varphi}:\left\Vert \iota\right\Vert \leq\sqrt{\frac{\lambda_{\varphi}\delta}{\lambda_{1}\lambda_{V}}}\right\} .\label{eq:equilibrium set}
\end{equation}

\begin{thm}
Consider the dynamical system described by (\ref{eq:Agent Dynamics})
and (\ref{eq:Target Dynamics}). For any initial conditions of the
state vector $z\left(t_{0}\right)\in\mathcal{S}$, the controller
given by (\ref{eq:control input}) and the adaptation law given by
(\ref{eq:Update Law}) ensure that $z$ uniformly exponentially converges
to $\mathcal{U}$ in the sense that {\footnotesize{}
\[
\left\Vert z\left(t\right)\right\Vert \leq\sqrt{\frac{\lambda_{\varphi}}{\lambda_{1}}\left\Vert z\left(t_{0}\right)\right\Vert ^{2}{\rm e}^{-\frac{2\lambda_{V}}{\lambda_{\varphi}}\left(t-t_{0}\right)}+\frac{\lambda_{\varphi}\delta}{\lambda_{1}\lambda_{V}}\left(1-{\rm e}^{-\frac{2\lambda_{V}}{\lambda_{\varphi}}\left(t-t_{0}\right)}\right)},
\]
} for all $t\in\mathbb{R}_{\geq t_{0}}$, provided that the sufficient
gain condition $k_{\min}>\lambda_{V}+\frac{1}{k_{2}}\rho\left(\sqrt{\frac{\lambda_{\varphi}\delta}{\lambda_{1}\lambda_{V}}}\right)$
is satisfied and Assumptions \ref{targetbounds} and \ref{thm:The-loss-function}
hold.
\end{thm}
\begin{IEEEproof}
Taking the total derivative of (\ref{eq:Lyapunov}) along the trajectories
of (\ref{eq:IVP}) yields{\footnotesize{}
\begin{equation}
\begin{aligned}\frac{{\rm d}}{{\rm d}t}V\left(z\left(t\right)\right) & =\nabla V\left(z\left(t\right)\right)^{\top}\frac{{\rm d}}{{\rm dt}}z\left(t\right)\\
 & =e^{\top}\left(t\right)\dot{e}\left(t\right)+r^{\top}\left(t\right)\dot{r}\left(t\right)+\widetilde{\Theta}^{\top}\left(t\right)\Gamma^{-1}\dot{\widetilde{\Theta}}\left(t\right).
\end{aligned}
\label{eq:vDot}
\end{equation}
}Substituting (\ref{eq:CLES}), invoking \cite[Appendix E.4]{Krstic.Kanellakopoulos.ea1995},
and using (\ref{eq: parameter estimation error}) yields{\small{}
\begin{equation}
\begin{aligned}\frac{{\rm d}}{{\rm d}t}V\left(z\left(t\right)\right) & \leq-k_{1}\left\Vert e\left(t\right)\right\Vert ^{2}-k_{2}\left\Vert r\left(t\right)\right\Vert ^{2}-k_{3}\left\Vert \widetilde{\Theta}\left(t\right)\right\Vert ^{2}\\
 & +r^{\top}\left(t\right)\left(R\left(\kappa\left(t\right),\widetilde{\Theta}\left(t\right)\right)+\varepsilon\left(\kappa\left(t\right)\right)\right)\\
 & -r^{\top}\left(t\right)\omega\left(t\right)+k_{3}\widetilde{\Theta}^{\top}\left(t\right)\Theta^{\ast}.
\end{aligned}
\label{eq:vDot-1}
\end{equation}
}Using (\ref{eq:Tracking Error}), Assumption \ref{targetbounds},
and the triangle inequality yields $\left\Vert q\left(t\right)\right\Vert \leq\left\Vert e\left(t\right)\right\Vert +\overline{q}_{d}$.
Similarly, using (\ref{eq: Filtered Tracking Error}), Assumption
\ref{targetbounds}, and the triangle inequality yields $\left\Vert \dot{q}\left(t\right)\right\Vert \leq k_{1}\left\Vert e\left(t\right)\right\Vert +\left\Vert r\left(t\right)\right\Vert +\overline{\dot{q}}_{d}$.
Therefore, using the definition of $\kappa$ yields
\begin{equation}
\left\Vert \kappa\left(t\right)\right\Vert \leq\left(k_{1}+2\right)\left\Vert z\left(t\right)\right\Vert +2\left(\overline{q}_{d}+\overline{\dot{q}}_{d}\right).\label{eq:kappaBounds}
\end{equation}
From \cite[Theorem 1]{Patil.Fallin.ea2025}, there exists a polynomial
$\rho_{0}\left(\left\Vert \kappa\right\Vert \right)=a_{2}\left\Vert \kappa\right\Vert ^{2}+a_{1}\left\Vert \kappa\right\Vert +a_{0}$
with $a_{2},a_{1},a_{0}\in\mathbb{R}_{\geq0}$ such that $\left\Vert R\left(\kappa,\widetilde{\Theta}\right)\right\Vert \leq\rho_{0}\left(\left\Vert \kappa\right\Vert \right)\left\Vert \widetilde{\Theta}\right\Vert ^{2}$.
Thus, using (\ref{eq:theta star}), (\ref{eq: reconstruction error bound}),
(\ref{eq: parameter estimation error}), (\ref{eq:kappaBounds}),
and the definition of $\overline{\Theta}$ yields $\left\Vert R\left(\kappa\left(t\right),\widetilde{\Theta}\left(t\right)\right)+\varepsilon\left(\kappa\left(t\right)\right)\right\Vert \leq2\overline{\Theta}\rho_{0}\left(\left(k_{1}+2\right)\left\Vert z\left(t\right)\right\Vert +2\left(\overline{q}_{d}+\overline{\dot{q}}_{d}\right)\right)\left\Vert \widetilde{\Theta}\left(t\right)\right\Vert +\overline{\varepsilon}$.

Since $\rho_{0}$ is a polynomial with non-negative coefficients,
it is strictly increasing. Consequently, there exists a strictly increasing
function $\rho:\mathbb{R}_{\geq0}\to\mathbb{R}_{\geq0}$ such that
$\rho\left(\left\Vert z\right\Vert \right)=3\overline{\Theta}^{2}\rho_{0}^{2}\left(\left(k_{1}+2\right)\left\Vert z\right\Vert +2\left(\overline{q}_{d}+\overline{\dot{q}}_{d}\right)\right)$.
Thus, using Young's inequality and the definitions of $\delta$ and
$k_{\min}$ yields that (\ref{eq:vDot-1}) is upper bounded as
\begin{equation}
\frac{{\rm d}}{{\rm d}t}V\left(z\left(t\right)\right)\leq-\left(k_{\min}-\frac{\rho\left(\left\Vert z\left(t\right)\right\Vert \right)}{k_{2}}\right)\left\Vert z\left(t\right)\right\Vert ^{2}+\delta.\label{eq:vDot-2}
\end{equation}

To establish the existence of a solution on $\mathbb{R}_{\geq t_{0}}$,
a contradiction argument is employed. Let $\left[t_{0},T_{\max}\right)$
be the maximal interval of existence for solution $t\mapsto z\left(t\right)$
to (\ref{eq:IVP}) with $z\left(t_{0}\right)\in\mathcal{S}$. Suppose,
for contradiction, that $T_{\max}<\infty$. Then $\lim_{t\to T_{\max}^{-}}\left\Vert z\left(t\right)\right\Vert =\infty$.

Define the set $\mathcal{I}\triangleq\left\{ t\right.\in\left[t_{0},T_{\max}\right):z\left(\tau\right)\in\mathcal{D}\text{ for all }\tau\in\left.\left[t_{0},t\right]\right\} $.
Since $z\left(t_{0}\right)\in\mathcal{S}\subset\mathcal{D}$ and the
solution $t\mapsto z\left(t\right)$ is continuous, $\mathcal{I}$
is non-empty and contains a non-trivial interval $\left[t_{0},t_{0}+\eta\right)$
for some $\eta>0$. Consequently, using (\ref{eq: RQ}) yields that
(\ref{eq:vDot-2}) is upper bounded as
\begin{align}
\frac{{\rm d}}{{\rm d}t}V\left(z\left(t\right)\right) & \leq-\frac{2\lambda_{V}}{\lambda_{\varphi}}V\left(z\left(t\right)\right)+\delta,\label{eq:vdot}
\end{align}
for all $t\in\mathcal{I}$. Solving the differential inequality given
by (\ref{eq:vdot}) over $\mathcal{I}$ yields{\scriptsize{}
\begin{equation}
V\left(z\left(t\right)\right)\leq V\left(z\left(t_{0}\right)\right){\rm e}^{-\frac{2\lambda_{V}}{\lambda_{\varphi}}\left(t-t_{0}\right)}+\frac{\lambda_{\varphi}\delta}{2\lambda_{V}}\left(1-{\rm e}^{-\frac{2\lambda_{V}}{\lambda_{\varphi}}\left(t-t_{0}\right)}\right),\label{eq:vsol}
\end{equation}
}for all $t\in\mathcal{I}$. Applying (\ref{eq: RQ}) to (\ref{eq:vsol})
yields{\tiny{}
\begin{equation}
\left\Vert z\left(t\right)\right\Vert \leq\sqrt{\frac{\lambda_{\varphi}}{\lambda_{1}}\left\Vert z\left(t_{0}\right)\right\Vert ^{2}{\rm e}^{-\frac{2\lambda_{V}}{\lambda_{\varphi}}\left(t-t_{0}\right)}+\frac{\lambda_{\varphi}\delta}{\lambda_{1}\lambda_{V}}\left(1-{\rm e}^{-\frac{2\lambda_{V}}{\lambda_{\varphi}}\left(t-t_{0}\right)}\right)},\label{eq:solution}
\end{equation}
}for all $t\in\mathcal{I}$. Since $z\left(t_{0}\right)\in\mathcal{S}$,
$\left\Vert z\left(t_{0}\right)\right\Vert \leq\sqrt{\frac{\lambda_{1}}{\lambda_{\varphi}}}\overline{\rho}^{-1}\left(k_{2}\left(k_{\min}-\lambda_{V}\right)-\rho\left(0\right)\right)-\sqrt{\frac{\delta}{\lambda_{V}}}$.
Substituting this into (\ref{eq:solution}) and using the fact that
${\rm e}^{-\frac{2\lambda_{V}}{\lambda_{\varphi}}\left(t-t_{0}\right)}\leq1$
for all $t\geq t_{0}$ yields $\left\Vert z\left(t\right)\right\Vert <\overline{\rho}^{-1}\left(k_{2}\left(k_{\min}-\lambda_{V}\right)-\rho\left(0\right)\right)$
for all $t\in\mathcal{I}$. This implies that $z\left(t\right)\in\text{int}\left(\mathcal{D}\right)$
for all $t\in\mathcal{I}$. By continuity of $z\left(t\right)$, if
$\sup\mathcal{I}<T_{\max}$, then $z\left(\sup\mathcal{I}\right)\in\partial\mathcal{D}$,
which contradicts the established bound. Therefore, $\sup\mathcal{I}=T_{\max}$
and thus $\mathcal{I}=\left[t_{0},T_{\max}\right)$.

Because the projection operator defined in defined in (\ref{eq:Update Law})
is locally Lipschitz \cite[Lemma E.1]{Krstic.Kanellakopoulos.ea1995},
the right-hand side of (\ref{eq:IVP}) is piecewise continuous in
$t$ and locally Lipschitz in $z$ for all $t\geq t_{0}$ and all
$z\in\mathbb{R}^{\varphi}$. Since $z\left(t\right)$ remains in the
compact set $\mathcal{D}$ for all $t\in\left[t_{0},T_{\max}\right)$,
the solution is uniformly bounded, meaning $\sup_{t\in\left[t_{0},T_{\max}\right)}\left\Vert z\left(t\right)\right\Vert <\infty$.
Therefore, by \cite[Theorem 3.3]{Khalil2002}, the solution can be
extended beyond $T_{\max}$, contradicting the maximality of $\left[t_{0},T_{\max}\right)$.
Therefore, $T_{\max}=\infty$, and the solution exists for all $t\in\mathbb{R}_{\geq t_{0}}$
with $z\left(t\right)\in\mathcal{D}$ for all $t\in\mathbb{R}_{\geq t_{0}}$.

Consequently, all trajectories with $z\left(t_{0}\right)\in\mathcal{S}$
satisfy (\ref{eq:solution}), for all $t\in\mathbb{R}_{\geq t_{0}}$.
As $t\to\infty$, this bound converges to $\left\Vert z\left(t\right)\right\Vert \leq\sqrt{\frac{\lambda_{\varphi}\delta}{\lambda_{1}\lambda_{V}}}$
which, by (\ref{eq:equilibrium set}), implies that the trajectory
$z\left(t\right)$ converges to $\mathcal{U}$. Furthermore, since
$z\left(t\right)\in\mathcal{D}$ for all $t\in\mathbb{R}_{\geq t_{0}}$,
it follows from (\ref{eq:OmegaSet}) and (\ref{eq:kappaBounds}) that
$\kappa\left(t\right)\in\Omega$ for all $t\in\mathbb{R}_{\geq t_{0}}$,
ensuring that the universal approximation property of the ResNet expressed
in (\ref{eq:UFAP}) holds for all time.

Because $\lambda_{V}$ is independent of the initial time $t_{0}$
or initial condition $z\left(t_{0}\right)$, the exponential convergence
is uniform \cite{Loria.Panteley2002}. Additionally, the boundedness
of $\left\Vert z\left(t\right)\right\Vert $ implies that $\left\Vert e\left(t\right)\right\Vert $,
$\left\Vert r\left(t\right)\right\Vert $, and $\left\Vert \widetilde{\Theta}\left(t\right)\right\Vert $
are bounded for all $t\in\mathbb{R}_{\geq t_{0}}$. Therefore, since
$q_{d},\dot{q}_{d}\in\mathcal{L}_{\infty}\left(\mathbb{R}_{\geq0};\mathbb{R}^{n}\right)$
by Assumption \ref{targetbounds}, using (\ref{eq:Tracking Error})
and (\ref{eq: Filtered Tracking Error}) yields that $q,\dot{q}\in\mathcal{L}_{\infty}\left(\mathbb{R}_{\geq0};\mathbb{R}^{n}\right)$,
implying that $g^{+}\left(q,\dot{q},t\right)$ is bounded. Following
(\ref{eq:kappaBounds}) and the fact that $z\in\mathcal{L}_{\infty}\left(\mathbb{R}_{\geq t_{0}};\mathbb{R}^{\varphi}\right)$
yields that $\kappa\in\mathcal{L}_{\infty}\left(\mathbb{R}_{\geq t_{0}};\mathbb{R}^{4n}\right)$.
Due to the projection operator, $\widehat{\Theta}\in\mathcal{L}_{\infty}\left(\mathbb{R}_{\geq t_{0}};\mathbb{R}^{{\tt p}}\right)$.
Since $\left(\kappa,\widehat{\Theta}\right)\in\mathcal{L}_{\infty}\left(\mathbb{R}_{\geq0};\mathbb{R}^{4n}\times\mathbb{R}^{{\tt p}}\right)$,
$\Psi\left(\kappa\left(t\right),\widehat{\Theta}\left(t\right)\right)$
is bounded. Thus, by (\ref{eq:control input}), $u$ is bounded.
\end{IEEEproof}

\section{Experiment}

\begin{table}
\centering{}\caption{\label{tab:simulation_parameters}Parameters used in the comparative
experiment of SNN, DNN, and ResNet -based adaptive controllers.}
\begin{tabular}{|c|c|c|c|}
\hline 
 & SNN & DNN & ResNet\tabularnewline
\hline 
Neurons & 8 & 2 & 2\tabularnewline
\hline 
Layers & 1 & 32 & 2\tabularnewline
\hline 
Blocks & 0 & 0 & 4\tabularnewline
\hline 
Parameters & \multicolumn{2}{c|}{254} & 226\tabularnewline
\hline 
Outer Activation & \multicolumn{3}{c|}{tanh}\tabularnewline
\hline 
Inner Activation & N/A & \multicolumn{2}{c|}{Swish\tablefootnote{$\text{Swish}\left(x\right)\triangleq x\cdot\sigma\left(x\right)$,
where $\sigma\left(x\right)$ is the sigmoid function \cite{Ramachandran.Zoph.ea2017}.}}\tabularnewline
\hline 
Shortcut Activation & \multicolumn{2}{c|}{N/A} & Swish\tabularnewline
\hline 
Learning Rate & $\Gamma=0.05$ & $\Gamma=0.1$ & $\Gamma=0.025$\tabularnewline
\hline 
Search Space Bound & $\overline{\Theta}=4$ & $\overline{\Theta}=8$ & $\overline{\Theta}=1$\tabularnewline
\hline 
Control Gains & \multicolumn{3}{c|}{$k_{1}=0.77$, $k_{2}=0.66$, $k_{3}=1{\rm e}^{-6}$}\tabularnewline
\hline 
\end{tabular}
\end{table}
Experimental validation was performed on a Freefly Astro quadrotor
equipped with a PX4 flight controller at the University of Florida's
Autonomy Park outdoor facility. State estimation utilized the onboard
EKF2 fusing GPS, optical flow, and Lidar data. A cascaded control
architecture was employed, where the proposed ResNet-based controller,
implemented as a 50 Hz ROS2 outer loop, generated acceleration commands
sent via MAVROS to the PX4's inner-loop controller operating at 400
Hz. The quadrotor autonomously tracked a 15 m \texttimes{} 5 m figure-eight
trajectory at 2.5 m altitude for 360 s. 
\begin{figure}
\begin{centering}
\includegraphics[width=1\columnwidth]{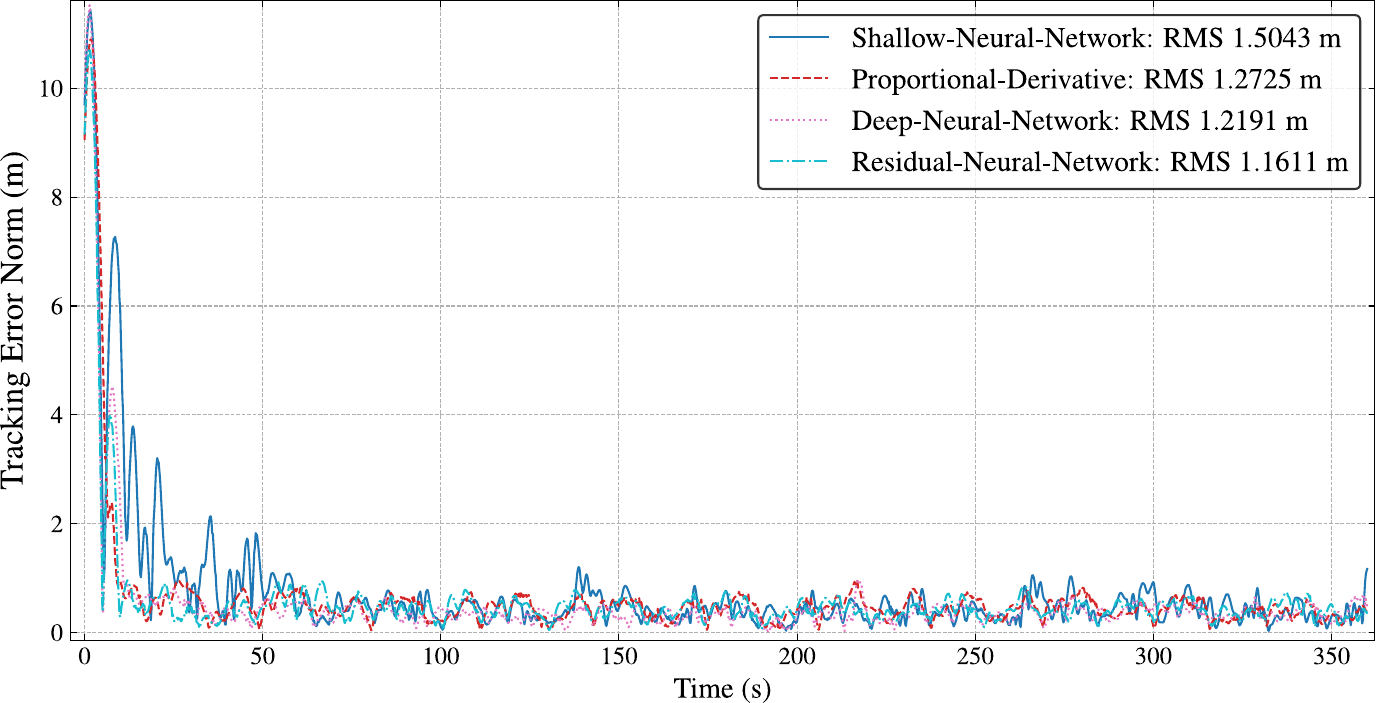}
\par\end{centering}
\caption{\label{fig:Tracking Error-1} Tracking error comparison over the 360-second
experiment.}
\end{figure}
The proposed controller was compared against three benchmarks: a proportional-derivative
(PD) controller, a shallow NN-based adaptive controller (SNN) and
a DNN-based adaptive controller (employing $\Phi$ instead of $\Psi$
in (\ref{eq:control input})). Controller parameters are provided
in Table \ref{tab:simulation_parameters}. Figure \ref{fig:Tracking Error-1}
illustrates that the proposed ResNet controller reduced tracking error
by 22.81\%, 8.75\%, and 4.76\% relative to the SNN, PD, and DNN controllers,
respectively, while using approximately 11\% fewer parameters compared
to the SNN and DNN controllers.
\begin{figure}
\begin{centering}
\includegraphics[width=1\columnwidth]{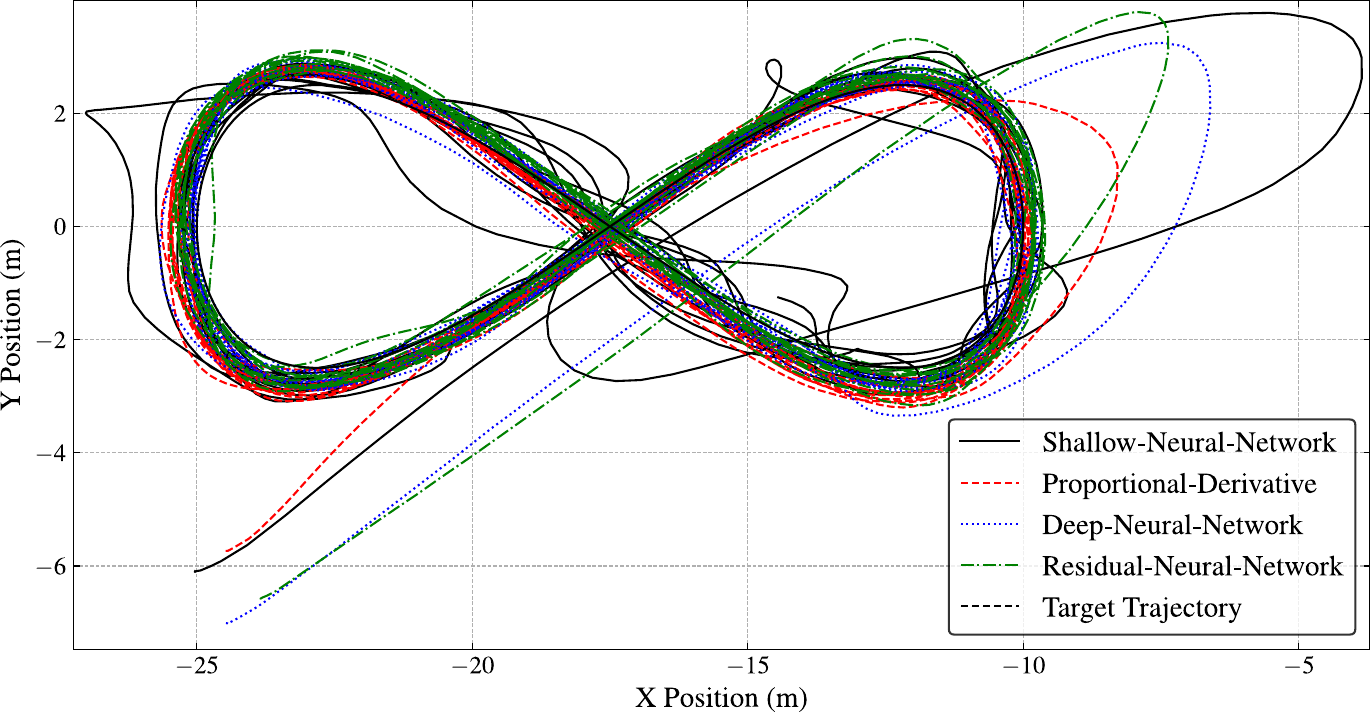}
\par\end{centering}
\caption{\label{fig:Tracking Error-1-1} Trajectory comparison over the 360-second
experiment. }
\end{figure}

\section{Conclusion}

The presented generalized ResNet architecture addresses adaptive control
of nonlinear systems with black box uncertainties. Key innovations
include pre-activation shortcut connections that improve signal propagation
and a zeroth layer block that allows handling different input-output
dimensions. The Lyapunov-based adaptation law guarantees exponential
convergence to a neighborhood of the target state.

\bibliographystyle{ieeetr}
\bibliography{Sources}

\end{document}